\documentclass{raa}

\usepackage{graphicx,times}             
\input{epsf.sty}
\input{psfig.sty}

\newcommand{\gsim}{\raisebox{-0.3ex}{\mbox{$\stackrel{>}{_\sim} \,$}}}
\newcommand{\lsim}{\raisebox{-0.3ex}{\mbox{$\stackrel{<}{_\sim} \,$}}}

\begin{document}


\title{Possible origin of viscosity in the Keplerian accretion disks due to secondary perturbation:
Turbulent transport without magnetic field}

\volnopage{Vol.0 (200x) No.0, 000--000}      

\author{Banibrata Mukhopadhyay \and   Kanak Saha 
}
\institute{Astronomy and Astrophysics Program, Department of Physics,
Indian Institute of Science, Bangalore 560012, India;
{\it bm@physics.iisc.ernet.in} \\
}

\date{Received~~2009 month day; accepted~~2009~~month day}

\abstract
{
The origin of hydrodynamic turbulence in rotating shear flow 
is a long standing puzzle. Resolving it is especially important in astrophysics when
the flow angular momentum profile is Keplerian which forms 
an accretion disk having negligible molecular viscosity. 
Hence, any viscosity in such systems must be due to turbulence, arguably
governed by magnetorotational instability especially when temperature $T\gsim 10^5$.
However, such disks
around quiescent cataclysmic variables, protoplanetary and star-forming disks,
the outer regions of disks in active galactic nuclei are practically neutral in charge
because of their low temperature, and thus expected not to be coupled with the magnetic field
appropriately to generate any transport due to 
the magnetorotational instability.
This flow is similar to plane Couette flow including the Coriolis force,
at least locally. What drives their turbulence and then transport, 
when such flows do not exhibit any unstable mode under linear hydrodynamic perturbation?
We demonstrate that the threedimensional secondary disturbance to the primarily
perturbed flow triggering elliptical instability may generate significant turbulent 
viscosity ranging $0.0001\lsim\nu_t\lsim0.1$ to explain transport in accretion flows.
}

\authorrunning{Mukhopadhyay \& Saha}

\titlerunning{Origin of viscosity in accretion disks due to secondary perturbation}

\maketitle

\section{Introduction}

One of the main problems behind the origin of hydrodynamic turbulence in shear flow 
is that there is a significant mismatch between the
predictions of linear theory and experimental data. For example,
in the case of plane Couette flow, laboratory experiments and
numerical simulations show that the flow may be turbulent at a
Reynolds number as low as $Re\sim 350$, while according to the linear
theory the flow should be stable for all $Re$. Similar mismatch
between theoretical results and observations is found in
astrophysical contexts, where the accretion flow
of neutral gas with Keplerian angular momentum profile, which essentially behaves like 
rotating shear flow, is a common subject. 
Examples of such flow systems are accretion disks
around quiescent cataclysmic variables (\cite{gm}),
protoplanetary and star-forming disks (\cite{blb}), and
the outer regions of disks in active galactic nuclei (\cite{mq}).

A Keplerian accretion disk flow having a very low
molecular viscosity must generate turbulence and successively
diffusive viscosity, which support the transfer of mass inwards and
angular momentum outwards. However, theoretically this
flow, in absence of magnetic field, never exhibits any unstable mode which could trigger
turbulence in the system. On the other hand, the laboratory
experiments of Taylor-Couette systems, which are similar to
Keplerian disks, seem to indicate that although the Coriolis force
delays the onset of turbulence, the flow is ultimately unstable
to turbulence for Reynolds numbers larger than a few thousand
(\cite{richard2001}), even for subcritical systems. 
Indeed, Bech \& Anderson
(1997) see turbulence persisting in numerical simulations of
subcritical rotating flows for large enough Reynolds numbers.

How does shearing flow that is linearly stable to perturbations
switch to a turbulent state?  Since last decade, many authors
including ourselves have come forward with a possible explanation of this fact 
based on {\it bypass} transition (see,
\cite{butfar,rh,tref,chagelish,umurhan,man} and references
therein) where the decaying linear modes show an arbitrarily
large transient energy growth at a suitably tuned perturbation.
In lieu of linear instabilities e.g. magnetorotational
instability, the transient energy growth, supplemented by a
non-linear feedback process to repopulate the growing
disturbance, could plausibly sustain turbulence for large enough
Reynolds numbers.

The behavior of shear flows, however, in the presence of rotation is enormously different 
compared to that in absence of rotation. The Coriolis
effect is the main
culprit behind this change in behavior killing any growth of energy even of transient kind
in the presence of rotation. In the case of shear flow with a varying angular velocity profile, e.g. 
Keplerian accretion flow, the above mentioned transient energy growth is
insignificant for threedimensional perturbations. 
To overcome this limitation, it is necessary to invoke additional effects.
Various kinds of secondary instability, such as the elliptical
instability, are widely discussed as a possible route to
self-sustained turbulence in linearly perturbed shear flows (see,
e.g. \cite{pier86,bay86,cc86,ls87,helor88,wal90,
c89,lrm96,ker02}). These effects, which generate
threedimensional instabilities of a twodimensional flow with
elliptical streamlines, have been proposed as generic mechanism
for the breakdown of many twodimensional high Reynolds number
flows whose vortex structures can be locally seen as elliptical
streamlines. Recently, one of the
present authors has studied the secondary perturbation
and corresponding elliptical vortex effects in accretion disks and
pinpointed that they can be the seed of threedimensional hydrodynamic instability (\cite{mukh06}).
Subsequently, by numerical simulation, this has been shown to be one of the
possible sources to generate turbulence to 
form large objects from the dusty gas surrounding a young star (\cite{cuzzi1,cuzzi2}).
Moreover, vortex generation in the unmagnetized protoplanetary disks has been
furnished by hydrodynamic turbulence (\cite{borro}) which leads to planet formation, and 
angular momentum transport in disks.
However, whether they lead to non-linear feedback and threedimensional turbulence
are yet to be shown explicitly.

Here we plan to show in detail that threedimensional secondary
perturbation generating large growth in the flow time scale may generate
significant turbulent viscosity
in rotating shear flows, more precisely in plane shear flows with the Coriolis force. 
The plane shear flow with the Coriolis force essentially behaves as a local patch
of a rotating shear flow. Possibility of significant turbulent transport in such flows 
by threedimensional perturbation opens a new window to explain
accretion process in flows which are neutral in charge. In particular, we address
the issue of deriving turbulent viscosity and the Shakura-Sunyaev viscosity parameter $\alpha$ (\cite{ss73})
from a pure hydrodynamical perspective \footnote{A preliminary calculation of such $\alpha$ has been
appeared in a collected volume of Gravity Research Foundation (\cite{bmij}).}. 
This is important for understanding
accretion flows in cold charge neutral medium.

It is important to note that transition to turbulence is not a unique process,
but it depends on the initial condition/disturbance and the nature
of the flow (\cite{sh01,crim03}). In fact, it is known that even in the presence of 
secondary instability, linearly unstable base flows may reach to a non-turbulent 
saturated state. However, turbulence definitely belongs to the nonlinear regime
and it is exhibited only in the situations when large growth of perturbation 
switches the system over the non-linear regime.
As our present goal is to understand the possible origin of hydrodynamic 
turbulence, we consider those situations when large energy growth governs 
non-linearity.

The paper is organized as follows. 
In the next section, we first recall the perturbation established previously (\cite{mukh06})
due to secondary disturbance in the Keplerian flow
and then discuss the range of corresponding Reynolds number and the solutions. Subsequently,
we estimate the corresponding turbulent viscosity of hydrodynamic origin in \S3.
We end in \S4 by discussing implications of our results.

\section{Perturbation and range of Reynolds number}

Considering a twodimensional velocity perturbation
$\vec{w}=(w_x(x,y,z,t),w_y(x,y,z,t),0)$,
and pressure perturbation
$p_p(x,y,z,t)$ in a small section of the Keplerian shear flow/disk, the linearized Navier-Stokes
and continuity equations for the incompressible fluid with plane background shear in the presence of a 
Coriolis component can be written 
in dimensionless units as (see \cite{man} for a detailed description)
\begin{equation}
{dw_x\over dt} = 2\Omega w_y - {\partial {p}_p\over \partial x}
+\frac{1}{Re} {\nabla}^2 w_x,
\label{xmmtm}
\end{equation}
\begin{equation}
{dw_y\over dt} = \Omega (q-2) w_x - {\partial {p}_p\over \partial y}
+\frac{1}{Re}{\nabla}^2 w_y,
\label{ymmtm}
\end{equation}
\begin{equation}
{\partial w_x\over\partial x} + {\partial w_y\over\partial y} 
= 0. 
\label{xmmec}
\end{equation} 
We consider the standard no-slip boundary condition such that $w_x=w_y=0$ at $x=\pm 1$ and 
according to the choice of variables in the coordinate system $\Omega=1/q$.
Here $(x,y,z)$ is a local Cartesian coordinate system centered at a 
point ($r,\phi)$ in the disk (\cite{man}) such that $dr=x$ and $rd\phi=y$.

When the Reynolds number is very large, the solution of eqns. (\ref{xmmtm}), 
(\ref{ymmtm}) and (\ref{xmmec}) are given by (\cite{man})
\begin{equation}
w_x=\zeta\frac{k_y}{l^2}\sin(k_xx+k_yy),\,\,w_y=-
\zeta\frac{k_x}{l^2}\sin(k_xx+k_yy)
\label{solpr}
\end{equation}
where $\zeta$ is the amplitude of vorticity perturbation, $k_x$ and $k_y$ are 
the components of primary perturbation wavevector and $l=\sqrt{k_x^2+k_y^2}$.
Under this {\it primary perturbation}, the flow velocity
and pressure modify to
\begin{eqnarray}
\vec{U}&=&\vec{U}^p+\vec{w}=(w_x,-x+w_y,0)={\bf A}.\vec{d},\,\,\,\bar{P}=\bar{p}+p_p,
\label{primper}
\end{eqnarray}
where $\vec{U}^p$, $\bar{p}$ are background velocity and pressure respectively, 
$\bf A$ is a tensor of rank $2$. 
Here $k_x=k_{x0}+k_yt$, which basically is the radial component of primary
perturbation wavevector, varying from $-\infty$ to a small number, where $k_{x0}$
is a large negative number: $|k_{x0}|\sim Re^{1/3}\sim t_{max}$ (\cite{man}). 

Now we concentrate on a further small patch of the primarily perturbed flow such that
the spatial scale is very small compared to the wavelength of primary 
perturbation satisfying $\sin(k_x x+k_y y)\sim k_x x=f\lsim 1$. 
In fact, $f\sim 1$ at close to the boundary of the patch when $y\rightarrow 0\,\,{\rm and}\,\,
2\pi/k_y$,
and at an intermediate location $f\ll 1$.  As $|k_x|$ varies from a large number to 
close to unity, the size of the primary perturbation box in the $x$-direction is 
$1/k_x\lsim 1$ when $k_y\sim1$, fixed. Hence, this further small patch must be confined
to a region: $-a\lsim x\lsim a$, when $f/|k_{x0}|\lsim a\lsim f$. 
Clearly, in this patch, $\vec{U}$ in eqn. (\ref{primper}) describes a flow having generalized
elliptical streamlines with $\epsilon=(k_x/l)^2$, a parameter related to the measure of
eccentricity \footnote{Note that $\epsilon$ is a parameter related to the measure
of eccentricity but not the eccentricity itself.},
running from $0$ to $1$ as the perturbation evolves. It was already shown (\cite{mukh06})
that a secondary perturbation in this background may grow exponentially leading
the flow unstable. We use this unstable flow in \S3, which was extensively discussed earlier 
(\cite{mukh06}), to derive $\nu_t$ and $\alpha$.

As we focus on the secondary perturbation at 
a small patch of the primarily perturbed shearing box, the variation of 
primary perturbation appears insignificant in the patch compared to that of the secondary one. 
Depending on the primary perturbation wavevector at a particular instant, the size of the 
secondary patch is appropriately adjusted. In fact $\epsilon$ varies
very very slowly and marginally deviates from unity 
in the time interval when $k_x$ varies from $k_{x0}$ (large 
negative) to, say, $-10$. Even when $k_x$ tends to $-3$, $\epsilon$ changes to $\sim 0.9$ only.
Therefore, $\epsilon$ and thus $\bf A$ practically remains constant. 

\subsection{Range of Reynolds number}

Due to consecutive choice of small boxes/patches, the Reynolds number in the secondary
flow is restricted with a particular choice of that in the primary flow.
Here in the interest of clarity, we work with the original dimensioned units.
The Reynolds number at the primary box is defined as
\begin{eqnarray}
Re_p=\frac{U_0 L}{\nu}=\frac{q\Omega_0 L^2}{\nu},
\label{pre}
\end{eqnarray}
where $2L$ is the box size in the $x$-direction and $2U_0$ is the relative velocity of the
fluid elements in the box between two walls along the $y$-direction. 
Now we recall the secondary perturbation at a smaller patch, extended from $-L_s$ to $+L_s$,
such that $|L_s|\sim aL$. To meet our requirement $\sin(k_x x+k_y y)\sim k_x x+k_y y$,
we remind that the small patch size needs to be adjusted.
Therefore, the Reynolds number at the secondary box is given by
\begin{eqnarray}
Re_s=\frac{q\Omega_0 L_s^2}{\nu}\sim\frac{q\Omega_0\,a^2 L^2}{\nu}.
\label{sre}
\end{eqnarray}
Hence,
\begin{eqnarray}
\frac{Re_p}{Re_s}\sim \frac{1}{a^2}\sim \frac{k_x^2}{f^2}.
\label{rre}
\end{eqnarray}
At the beginning of the primary perturbation $k_x=k_{x0}$ and thus $\epsilon=1$.
At this stage, the secondary box size $L_s=Lf/k_{x0}$ and $Re_p\gsim k_{x0}^2\,Re_s$.
With time $k_x$ decreases in magnitude but $\epsilon$ deviates little from unity until
$k_x\sim -3$ when $\epsilon=0.9$. Hence $\bf A$ can be considered constant approximately
as described above. At this stage $Re_p\ge 9\,Re_s$, atleast an order of
magnitude higher than $Re_s$. If the energy growth due to primary perturbation is maximized 
for $k_x=k_{x,min}=\pi$ (\cite{man}), then the range of $Re$ for the secondary
perturbation is given by $Re_p\,f^2/k_{x0}^2\lsim Re_s \lsim Re_p\,f^2/10$.
At $k_x=\pi$, $Re_s$ is atleast an order of magnitude lower than
$Re_p$. When $k_{x,mim}=1$, $Re_p\sim Re_s$ for $f\sim 1$. 
In general $Re_p\,f^2/k_{x0}^2\lsim Re_s \lsim Re_p\,f^2/k_{x,min}^2$.

\subsection{Solution}

Following previous work (\cite{mukh06}), the general solution for the evolution 
of secondary perturbation in the flow discussed above
can be written in terms of Floquet modes 
\begin{equation}
u_i(t)=\exp(\sigma\,t)\,f_i(\phi)\exp[i(k_1 x+k_2 y+k_3 z)],
\label{flo}
\end{equation}
where $\phi=\varpi\,t$, $f_i(\phi)$ is a periodic function having
time-period $T=2\pi/\varpi$, $\sigma$ is the Floquet exponent,
$k_1,k_2,k_3$ are the components of wavevector of the secondary perturbation.
Note that $\sigma$ is different at different $\epsilon$.
Clearly, if $\sigma$ is positive, then
the system is unstable. The detailed solutions were discussed elsewhere (\cite{mukh06})
what we will not repeat here.

In principle, $k_x$ varies with time and thus $\bf A$ does so.
Thus, generalizing the solution (\ref{flo}) for a (slowly) varying $\bf A$, we obtain
\begin{equation}
u_i(t)=\exp\left(\int\sigma(t)\,dt\right)f_i(\phi)\exp[i(k_1 x+k_2 y+k_3 z)],
\label{genv}
\end{equation}
where $\phi=\int\varpi(t)\,dt$. 
The eqns. (\ref{flo}) and (\ref{genv}) practically describe the solutions
for the entire parameter regime exhibiting elliptical vortices which are
very favorable for the elliptical instability to trigger.

For the present purpose, the physically interesting quantity is the energy 
growth of perturbation which is given by
\begin{equation}
G=\frac{|u_i(t)|^2}{|u_i(0)|^2}=\exp\left[2\,\Sigma(t)\right]
\frac{f_i^2(\phi)}{f_i^2(0)},
\label{grow}
\end{equation}
where $\Sigma(t)=\int\sigma(t)\,dt$ and $t=(k_x-k_{x0})/k_y$.
As $k_x(t)$ varies from a large negative value, $k_{x0}$,
to $0$, $t$ increases from $0$ to $t_{\rm max}=-k_{x0}/k_y$. Thus, 
the energy growth is controlled by the quantity $\Sigma(t)$, as
$f_i^2(\phi)/f_i^2(0)$ simply appears to be a phase factor. Therefore, our aim 
should be to evaluate $\Sigma$ for various possible perturbations.

\begin{figure}
\begin{center}
\includegraphics[width=18cm,height=18cm]{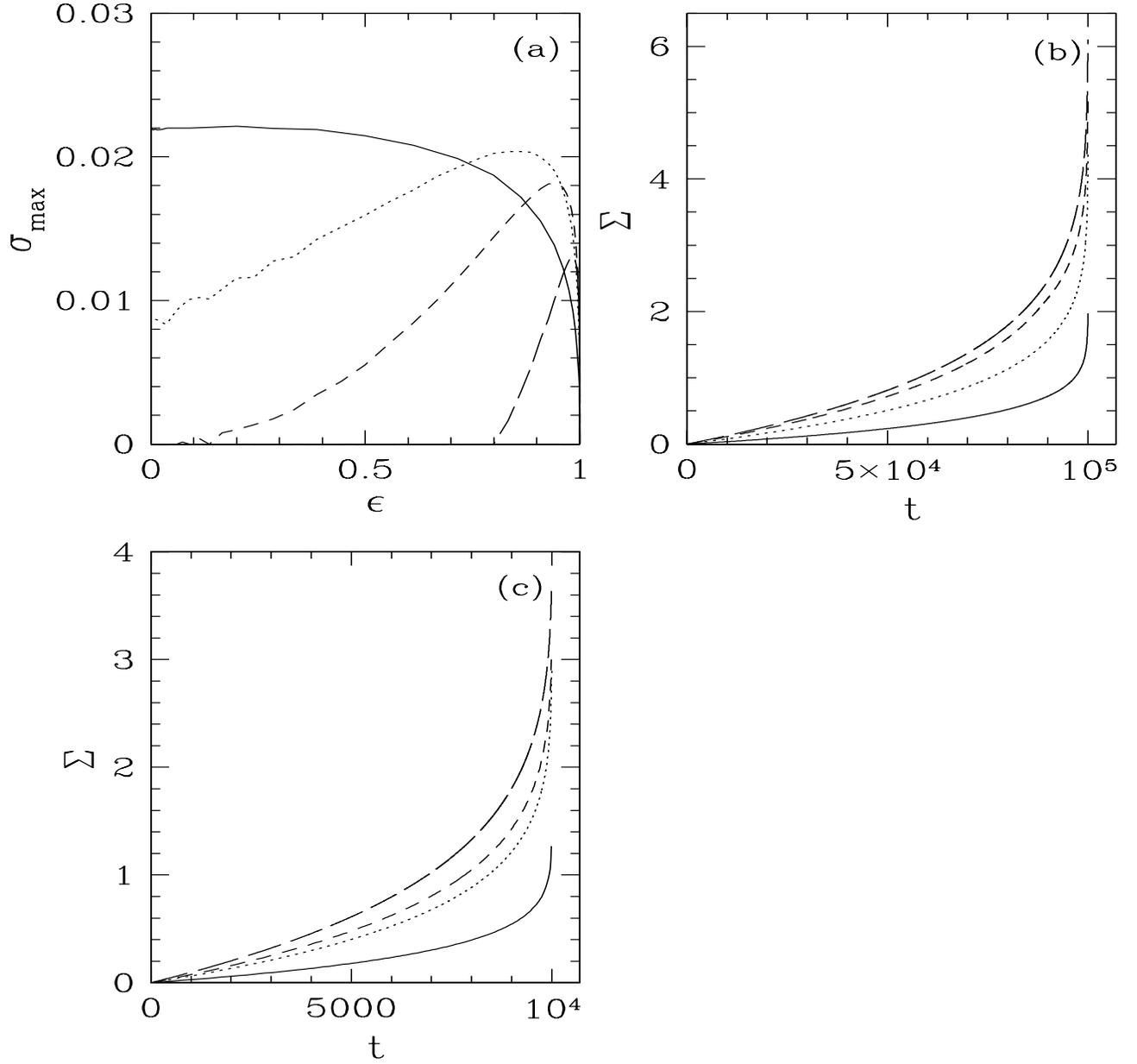}
\end{center}
\vskip2.5cm 
\caption{(a) Variation of maximum velocity growth rate as a
function of eccentricity parameter. 
Solid, dotted, dashed and
long-dashed curves indicate the results for $\zeta=0.01,0.05,0.1,0.2$ respectively (\cite{mukh06}). 
(b) Variation of $\Sigma$
as a function of time for $k_{x0}=-10^5$,
when various curves are same as of (a). (c) Same as (b) but for $k_{x0}=-10^4$. Other parameters are 
$k_{y0}=1$, $k_{10}=0$, $|\vec{k}_0|=1$, and $q=3/2$.
} \label{fig1}
\end{figure}

Let us specifically concentrate on the Keplerian accretion flows.
Figure \ref{fig1}a shows the variation of maximum velocity growth rate, $\sigma_{\rm max}$, as
a function of eccentricity parameter, $\epsilon$,
for the various choices of amplitude of vorticity, $\zeta$. By ``maximum'' we refer
the quantity obtained by maximizing over the vertical component of
the wavevector, $k_{3}$.
At large $\epsilon$ (as well as large $k_x$), when $\zeta$ is large, the background flow structure, 
$\bf A$, is elliptical with high eccentricity.
Therefore a vertical perturbation triggers the best growing mode
into the system.
However, with the decrease of $\zeta$, $\bf A$ approaches to that of the plane shear and
thus the growth rate decreases significantly.
At this stage, the corresponding best perturbation is
threedimensional but not the vertical one.

At small $\epsilon$ (and then small $k_x$), when $\zeta$ is large the eccentricity of the
background elliptical flow decreases significantly, and thus the
growth rate decreases. In this low eccentric flow, the best growth
rate arises due to the twodimensional perturbation. On the other
hand, when $\zeta$ is small, the background reduces to that of the plane shear flow.
Therefore, the growth rate
increases according to the shearing effects, as described by \cite{man}. An
interesting fact to note is that except the case of small $\epsilon$ ($k_x$) with a
large $\zeta$, the growth rate maximizes for the
threedimensional perturbation. Moreover, at a large $\zeta$ and a large 
$\epsilon$, the best growth rate arises due to a vertical (or almost vertical)
perturbation. 


As the accretion time scale is an important
factor, for the present purpose, physically interesting quantity is $\Sigma$ rather than $\sigma$ itself.
Figures \ref{fig1}b,c show the variation of $\Sigma$
as a function of $t$ at various $\zeta$. As the perturbation evolves with time, 
the corresponding $\Sigma$ increases. It is also clear that $\Sigma$ and then 
corresponding growth increases with the increase of $|k_{x0}|$ (and then $Re$), i.e. the increase
of accretion time scale, in addition to the increase of $\zeta$. In Table 1, we enlist the
approximate values of maximum growth factor, as follows from eqn. (\ref{grow}), corresponding
to $\Sigma_{\rm max}=\int_0^{t_{\rm max}}\sigma\,dt$, for the cases shown in Figs. \ref{fig1}b,c.
When $k_{x0}=-10^4$, $Re_p\sim 10^{12}$ (as $Re_p\sim t_{\rm max}^3\sim k_{x0}^3$) and
from eqn. (\ref{rre}) $Re_s(f=1)\gsim10^{4}$, the
maximum growth factor is significant for a
large amplitude of vorticity perturbation i.e. $\zeta>0.1$. However, the growth factor increases
with the increase of $Re_p$ and when $Re_p\sim 10^{15}$, and then $Re_s(f=1)\gsim 10^{5}$,
it is quite significant for an amplitude of vorticity perturbations
as small as $0.05$.
Therefore, it appears that a suitable threedimensional secondary perturbation
efficiently triggers elliptical instability and possible turbulence 
in rotating shear flows including accretion disks.

\centerline{\large Table 1} 
\centerline{\large Maximum energy growth corresponding
to cases shown in Figs. \ref{fig1}b,c}
\begin{center}
\begin{tabular}{cccc}
\hline
\hline
$|k_{x0}|$ & $\zeta$ & $\Sigma_{\rm max}$ & $G_{\rm max}$ \\
\hline
\hline
$10^5$ &$0.2$ & $6.1$ & $2\times 10^5$ \\
$10^5$ &$0.1$& $5.2$ & $3.3\times 10^4$ \\
$10^5$ &$0.05$& $4.43$ & $7\times 10^3$\\
$10^5$ &$0.01$& $1.97$ & $52$\\
\hline
\hline
$10^4$ &$0.2$ & $3.65$ & $1500$ \\
$10^4$ &$0.1$ & $3$ & $400$\\
$10^4$ &$0.05$& $2.9$ & $330$\\
$10^4$ &$0.01$& $1.27$ & $13$ \\
\hline
\hline
\end{tabular}
\end{center}

\section{Turbulent viscosity}

Here we attempt to quantify the turbulence by parametrizing it in terms of the viscosity.
This is essentially important, as explained in \S1, in flows like astrophysical accretion 
disks, where 
molecular viscosity is negligible, to explain any transport therein.

The tangential stress at a point $(r,\phi)$ of a rotating flow exhibiting turbulence is
\begin{eqnarray}
W_{r\phi}=\nu_t\, r \frac{d\Omega}{dr}=-\nu_t\, q \Omega,
\label{visdef}
\end{eqnarray}
where $\nu_t$ is the turbulent viscosity and $\Omega=\Omega_0 (r/r_0)^{-q}$. Note that
$q=3/2$ for the Keplerian angular velocity profile.
The perturbation described above is 
expected to govern the nonlinearity after certain time, say $t_g$. We also assume that the nonlinearity
leads to turbulence attributing the fact that at the initiation of turbulence the eddy velocity is 
same as the perturbation velocity. Therefore, we obtain the averaged tangential 
stress due to perturbation at $t=t_g$
\begin{eqnarray}
\nonumber
T_{r\phi}(t_g)\rightarrow T_{xy}(t_g)
=<u_x u_y>\\=\frac{k_2}{4\pi L_s}\int_{-L_s}^{+L_s}\int_0^{2\pi/k_2} u_x(t_g) u_y(t_g) dxdy,
\label{stress}
\end{eqnarray}
where we remind that the azimuthal flow is considered to be periodic in $y=2\pi/k_2$.


Now combining eqns. (\ref{visdef}), (\ref{stress}) and after some algebra we obtain
\begin{eqnarray}
\bar{\nu}_t=-\frac{T_{xy}}{q\Omega\left(\frac{h}{r}\right)M}
\label{alf}
\end{eqnarray}
where $T_{xy}=\int W_{xy}\,dxdy$, $M=\Omega x/c_s$ and $\bar{\nu}_t$ denotes the averaged
$\nu_t$ in the small section, computed here at $t=t_g$.  

Without any proper knowledge of turbulence in Keplerian flows which arise in 
accretion disks, Shakura \& Sunyaev (\cite{ss73}) 
parametrized it by a constant $\alpha$ considering  $W_{r\phi}$
to be proportional to the sound speed, $c_s$, given by
\begin{eqnarray}
W_{r\phi}=-\alpha c_s^2.
\label{ss}
\end{eqnarray}
$\alpha$ is called the Shakura-Sunyaev viscosity parameter.
They assumed that the small section under consideration to be isotropic so that
scaled the characteristic length $l_t$ of turbulence 
in terms of the largest macroscopic length scale of the disk, 
i.e. half-thickness $h$, and the eddy  velocity of turbulence $v_t$ in terms of sound speed $c_s$. Thus 
they defined the turbulent viscosity 
\begin{eqnarray}
\nu_t=\frac{l_t\,v_t}{3}=\alpha c_s h,
\label{turvis}
\end{eqnarray}
where $l_t=\alpha_l h$, $v_t=\alpha_v c_s$, $\alpha=\alpha_l\alpha_v/3$.
Obviously $\alpha_l\le 1$. If the turbulent velocity becomes supersonic,
then shock forms and reduces the velocity below the sound velocity
which assures $\alpha_v\le 1$. Therefore, $\alpha \lsim 1$.
From eqns. (\ref{alf}) and (\ref{turvis}) we write
\begin{eqnarray}
\bar{\alpha}=-\frac{T_{xy}}{q\Omega^2\left(\frac{h}{r}\right)^3M r^2},
\label{alf2}
\end{eqnarray}
where $\bar{\alpha}$ denotes the averaged
$\alpha$ in the small section. Therefore, if we know the
structure of the flow, then we can compute the turbulent viscosity due to various perturbations.
As we consider the size of
the section to be very small, $\bar{\alpha}$ and $\bar{\nu}_t$ are effectively equivalent to $\alpha$
and $\nu_t$ at a particular position in the disk.
Below we compute $T_{xy}$ for the various secondary perturbations and the corresponding
turbulent viscosities, at least in certain approximations.

\subsection{Secondary perturbation evolves much rapidly than the primary one}

From eqn. (\ref{flo}) we can write the velocity perturbation components
\begin{eqnarray}
\nonumber
u_x(x,y)&=&A_x\,e^{\sigma t} f_x(\phi)\sin(k_1\,x+k_2\,y+k_3\,z),\\
u_y(x,y)&=&A_y\,e^{\sigma t} f_y(\phi)\sin(k_1\,x+k_2\,y+k_3\,z),
\label{3dp}
\end{eqnarray}
where $A_x$ and $A_y$ are the amplitudes of perturbation modes, 
$k_{10}, k_{20}$ are the radial and the azimuthal components respectively of 
the secondary perturbation wavevector at $t=0$, 
$A_x$ and $A_y$ can be evaluated by the condition that the velocity components of 
the secondary perturbation reduce to that of the primary perturbation at $t=0$ 
(at the beginning of the evolution of secondary perturbation) given by
\begin{eqnarray}
\nonumber
A_x&=&\zeta\frac{k_y}{l^2(\epsilon)}\frac{C}{f_x(0)},\,\,\,
A_y=-\zeta\frac{k_x(\epsilon)}{l^2(\epsilon)}\frac{C}{f_y(0)},\\
C&=&\frac{\sin(k_x(\epsilon)\,x+k_y\,y)}{\sin(k_{10}\,x+k_{20}\,y+k_{30}\,z)},
\label{3da}
\end{eqnarray}
where $k_x(\epsilon)=\sqrt{\epsilon/(1-\epsilon)}k_y$, $C$ is of the order of unity
(for details see \cite{man,mukh06}). Therefore, from eqn. (\ref{stress})
\begin{eqnarray}
\nonumber
T_{xy}(t_g)&\sim&-\zeta^2\frac{k_x(\epsilon)k_y}{2l^4(\epsilon)}\,e^{2\sigma t_g}\,D,\\
D&=&C^2\frac{f_x(\phi)f_y(\phi)}{f_x(0)f_y(0)}.
\label{3txy}
\end{eqnarray}
Now by considering a typical case with $k_y=0.71$, $\nu_t$ and $\alpha$
can be computed as functions $\epsilon$ ($k_x$), when we know the time of evolution
of the secondary perturbation $t_g$.

\begin{figure}
\begin{center}
\includegraphics[width=18cm,height=18cm]{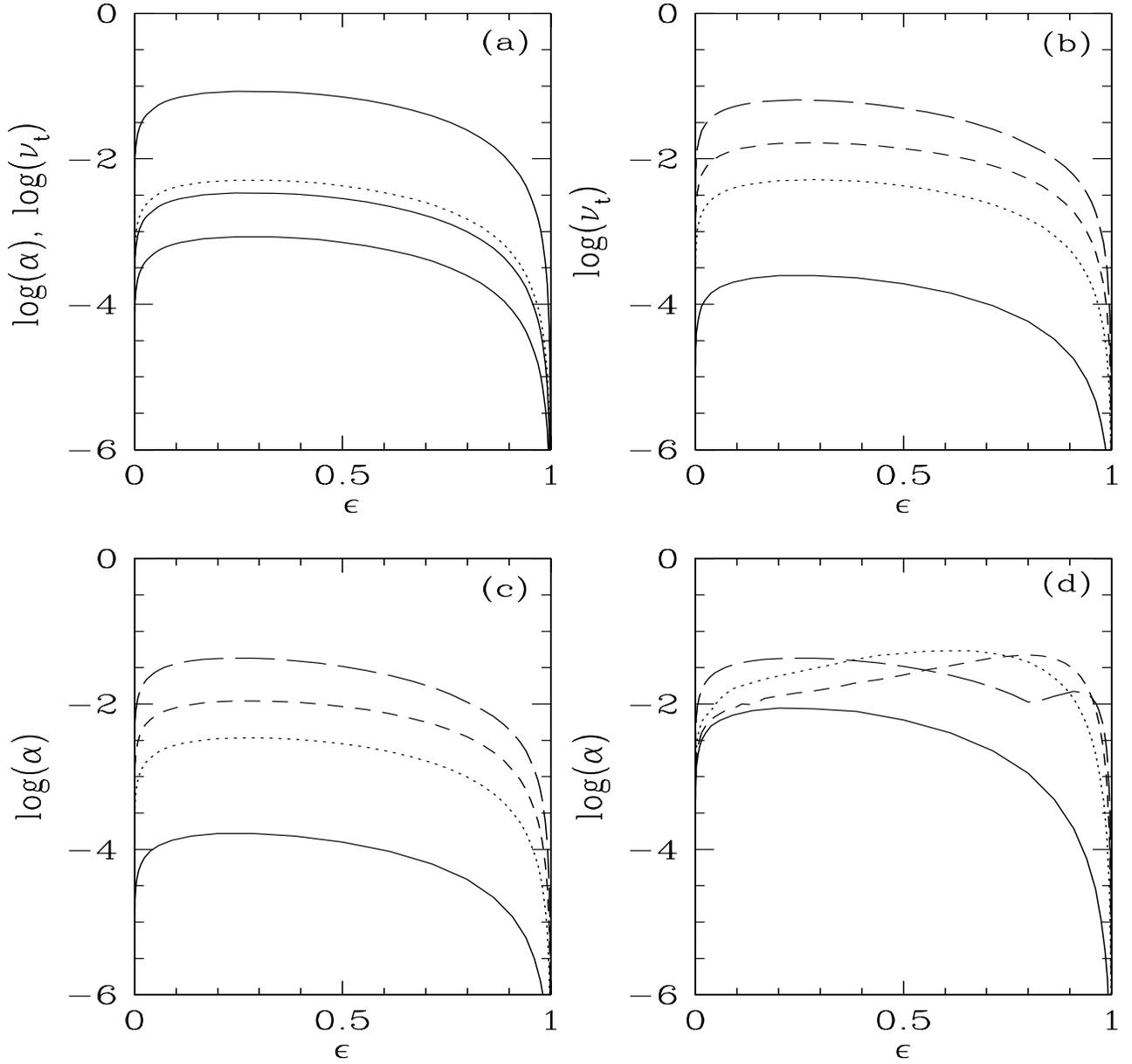}
\end{center}\vskip-1.6cm
\caption{
This is for the perturbation described in \S 3.A.
(a) Variation of $\nu_t$ (dotted curve) and $\alpha$ (solid curve) as functions of 
$\epsilon$ for $\zeta=0.05$ case described in Fig. \ref{fig1}a,
when $h(r)/r=0.01, 0.05, 0.1$
respectively for the top, middle, bottom curves of $\alpha$;
$r=30$, $k_y=0.71$, $t_g=10$. 
(b) Variation of $\nu_t$ as a function of
$\epsilon$ for the cases described in Fig. \ref{fig1}a with $h(r)/r=0.05,t_g=10,k_y=0.71$,
when solid, dotted, dashed, long-dashed curves correspond to
$\zeta=0.01,0.05,0.1,0.2$ respectively with $|\vec{k}_0|=1$.
(c) Same as in (b) except $\alpha$ is plotted in place of $\nu_t$.
(d) Same as in (c) except $t_g=100$.
  }
\label{fig2}
\end{figure}

Figure \ref{fig2} describes $\nu_t$ and $\alpha$ according to eqns. (\ref{alf}), (\ref{alf2}) and
(\ref{3txy}) for various disk parameters. As the primary perturbation evolves,
elliptical vortices form into the shearing flow which generate the turbulent viscosity
under a further perturbation. 
Figure \ref{fig2}a shows that the
viscosity varies with the eccentricity of vortices.
At a very early stage when the primary perturbation
is effectively a radial wave and $\epsilon\rightarrow 1$, the maximum velocity growth rate due
to secondary perturbation, $\sigma_{max}$ (shown in Fig. \ref{fig1}a), 
and the corresponding turbulent viscosity are very 
small, independent of the value of $\zeta$.
With time, the primary perturbation wavefronts
are straightened out by the shear until $t=t_{max}$, when the perturbation becomes
effectively an azimuthal wave and $\epsilon\rightarrow 0$. At this stage, $\sigma_{max}$
and the turbulent viscosity due to the secondary perturbation 
become zero again. This feature is clearly understood from eqn. (\ref{3txy}). However, at an 
intermediate time when $k_x(\epsilon)$ is finite, $\nu_t$ may be $\sim 0.005$ 
even in a moderately slim disk
with $h(r)/r=0.05$, when the time of evolution of secondary perturbation $t_g=10$. 
This $t_g$ is considered to be the time at which turbulence is triggered in the system.
Figures \ref{fig2}b-d show the variation of $\nu_t$ and $\alpha$ with the eccentricity of vortices 
at various $\zeta$ when $t_g=10,100$.
It is interesting to note, particularly for $t_g=100$, 
that with the increase of $\zeta$, first viscosity increases then
decreases. This is understood from the underlying energy growth rate shown in Fig. \ref{fig1}a,
when the readers are reminded that $\sigma=\sigma(\zeta,\epsilon)$. 
Note that the qualitative behavior of $\nu_t$ is same as that of $\alpha$. 
If we look at a typical case with $\zeta=0.05$ where
$\sigma=\sigma_{max}$ at $\epsilon=0.86$ which corresponds to $k_x=-1.76$, then $\alpha$ and $\nu_t$
computed at $t=t_g$ are for $Re_s \lsim Re_p\sim 10^8$.

\subsection{Secondary perturbation over the slowly varying primary perturbation}

In principle, the primary perturbation may vary with time during the evolution of secondary
perturbation. 
By numerical solutions, simultaneous evolution of the primary and the secondary perturbation along
with the corresponding energy growth has already been discussed earlier (\cite{mukh06}).
For the convenience of analytical computation of viscosity,
here we consider the regime of slow variation of the primary perturbation compared to the secondary one. 
Hence we recall eqn. (\ref{genv})  and
write the velocity perturbation components
\begin{eqnarray}
\nonumber
u_x&\rightarrow&u_{x_\Sigma}(x,y)=B_x\,e^{\Sigma(t)} f_x(\phi)\sin(k_1\,x+k_2\,y+k_3\,z),\\
u_y&\rightarrow&u_{y_\Sigma}(x,y)=B_y\,e^{\Sigma(t)} f_y(\phi)\sin(k_1\,x+k_2\,y+k_3\,z),
\label{3dps}
\end{eqnarray}
with $\phi=\int\varpi(t)dt$.
The amplitudes of perturbation modes $B_x$ and $B_y$
can be evaluated by the initial condition of secondary perturbation.
The secondary perturbation could trigger elliptical instability 
only after significant vortex forms in the flow due to the evolution of primary one.
At the beginning of the evolution of primary perturbation $k_{x0}\rightarrow-\infty$
(we choose the cases $k_{x0}=-10^5$ and $-10^4$) which corresponds to $\epsilon\rightarrow 1$
and thus effectively a plane shear background when
$\zeta$ is small (see \cite{mukh06}). In absence of vortex, this can 
not trigger elliptical instability under a secondary perturbation. As $k_{x0}$
decreases in magnitude, $\epsilon$ deviates from unity giving rise to a background
consisting of elliptical vortices. Above certain $\epsilon=\epsilon_c$, the secondary
perturbation does not have any effect to the primarily perturbed flow and $u_{x_\Sigma}$
and $u_{y_\Sigma}$ reduce to the primary perturbation. We hypothesize that $\epsilon_c=0.9999$.
Hence, $B_x$ and $B_y$ are computed in a similar fashion as in \S 3.A given by
\begin{eqnarray}
\nonumber
B_x&=&\zeta\frac{k_y}{l^2(\epsilon_c)}\frac{C}{f_x(0)},\,\,\,
B_y=-\zeta\frac{k_x(\epsilon_c)}{l^2(\epsilon_c)}\frac{C}{f_y(0)},\\
C&=&\frac{\sin(k_x(\epsilon_c)\,x+k_y\,y)}{\sin(k_{10}\,x+k_{20}\,y+k_{30}\,z)}.
\label{bxy}
\end{eqnarray}
Hence, from eqn. (\ref{stress}) the stress tensor  
\begin{eqnarray}
\nonumber
T_{xy}(t_{\rm max})&\sim&-\zeta^2\frac{k_x(\epsilon_c)k_y}{2l^4(\epsilon_c)}\,
e^{2\Sigma_{\rm max}}\,D,\\
D&=&C^2\frac{f_x(\phi)f_y(\phi)}{f_x(0)f_y(0)}
\label{txyps}
\end{eqnarray}
where $k_x$ reduces to zero at $t=t_{\rm max}$, which corresponds to the beginning of turbulence 
when $\Sigma=\Sigma_{\rm max}$.

\begin{figure}
\begin{center}
\includegraphics[width=\columnwidth]{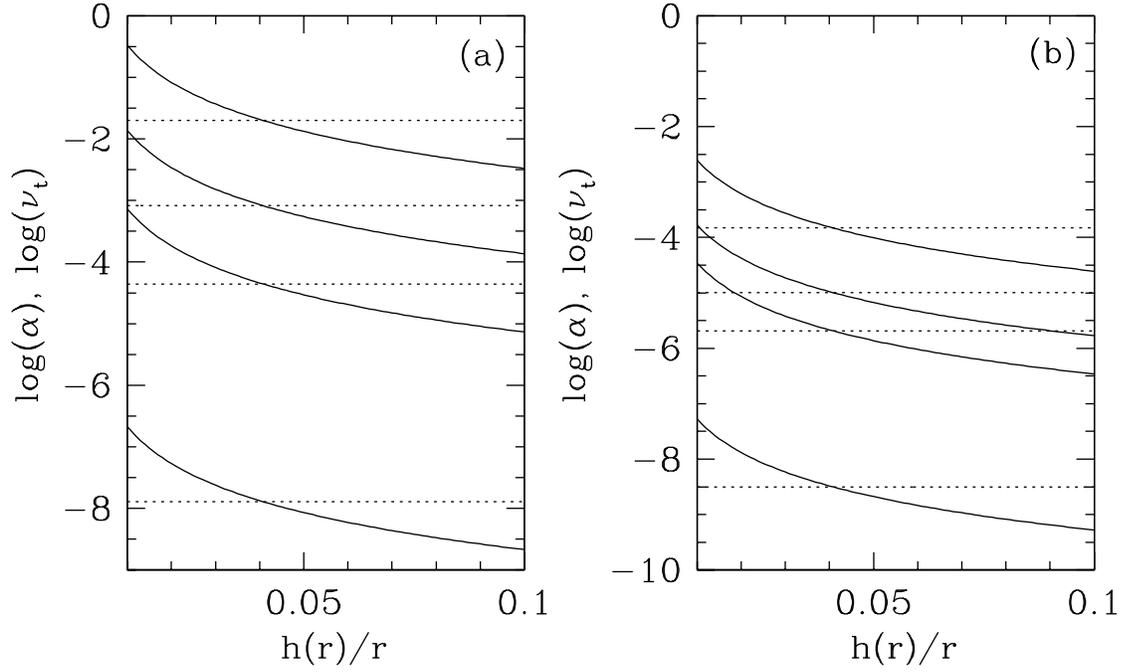}
\end{center}\vskip-3.6cm
\caption{
This is for the perturbation described in \S 3.B.
Variation of $\nu_t$ (dotted curve) and $\alpha$ (solid curve) as functions of $h(r)/r$
for cases shown in Figs. \ref{fig1}b,c, when
the curves from top to bottom correspond to $\zeta=0.2,0.1,0.05, 0.01$ with $r=30$ for
(a) $k_{x0}=-10^5$, (b) $k_{x0}=-10^4$. Other parameters are $k_y=1$, $\epsilon_c=0.9999$. 
  }
\label{fig3}
\end{figure}

It is found from Fig. \ref{fig3} that in a thin disk with $h(r)/r=0.01$, $\alpha$ at $r=30$ may be
as high as $\gsim 0.1$ for $k_{x0}=-10^5$ when $\zeta$ is very large. Although the viscosity 
decreases with the decrease of $\zeta$,
$\alpha$ still may be $\sim 0.001$ when $\zeta=0.05$. The turbulent viscosity decreases in a 
considerably thicker disk, but still $\alpha\sim 0.003$ at $h(r)/r=0.1$ when $\zeta=0.2$. 
For $\zeta\ge0.1$, $\nu_t\gsim0.001$ when $k_{x0}=-10^5$. The values of $\nu_t$ and $\alpha$ 
both decrease when $|k_{x0}|$ decreases to $10^4$, which is expected from Table 1 as well.
In this case, a significant turbulent viscosity
generates only at a large $\zeta=0.2$.

\section{Implications and Discussions}

Above results verify that at a range of $\epsilon$, 
the threedimensional growth rate due to
secondary perturbation in rotating shear flow of the Keplerian kind
is always real and positive and corresponding growth may be
exponential and significant enough, at least for a suitable choice of $\zeta$ and/or $Re$, 
to trigger non-linearity and then plausible turbulence in the flow time scale. 
With the increase of $k_{x0}$ ($\sim {Re_p}^{1/3}$),
the effect due to elliptical instability increases,
and thus corresponding growth does so.

As this growth is the result of threedimensional perturbation, 
underlying perturbation effect should survive even in the presence
of viscosity. 
There are many important natural phenomena where the Reynolds number is very large.
In astrophysical accretion disks, what applications are essentially considered in the present paper,
$Re$ always could be $\gsim10^{10}$ 
because of their very low molecular viscosity. Therefore, the
present mechanism is certainly applicable to 
such disk flows to resolve their {\it turbulence puzzle} when especially
it is cold
and neutral in charge and thus not a very plausible candidate for the magnetorotational instability. 
On the other hand, we suggest that the subcritical transition
to turbulence in Couette flow may be the result of secondary perturbation which
triggers elliptical instability modes into the system. 

We have tried to estimate the corresponding hydrodynamic turbulent viscosity. 
We have aimed to quantify the amount of turbulence through this
using the perturbations as the source of turbulence. We report here an observable range of 
viscosity obtained for the typical thin accretion disks and with reasonable values of flow vorticity. 
In place 
of $r=30$, if we choose the shearing box at a large distance from the central object, say at $r=500$,
then the computed $\alpha$ naturally decreases three orders of magnitude [see eqn. (\ref{alf2})]. 
We show by an extensive 
analysis the 
dependence of viscosity on the aspect ratio $(h/r)$ of the flow. The values of $\nu_t$ and 
$\alpha$ increase quite 
rapidly as the disk becomes thin to thinner.
From eqns. (\ref{alf}) and (\ref{alf2})
and with the results given in Figs. \ref{fig2} and \ref{fig3}, 
we find that it still might be as large as $10^{-4}$ for a thin disk 
even at a large distance, say, $r=500$. 

While some earlier laboratory experiments 
(e.g. \cite{richard2001}) predicted sub-critical transition 
to turbulence and then transport in hydrodynamical shear flows like
accretion disks, experiments by Ji et al. (2006) have argued
against it. Non-detection of turbulence and then any angular momentum 
transport of purely hydrodynamic origin could be due to the following facts. 
Maximum Reynolds number in this experiment is $2 \times 10^{6}$ whereas the cold disks such as
the protoplanetary disks have Reynolds number $\sim 10^{12}$. 
However, the critical Reynolds number for these systems could be $\sim 10^6-10^7$ or more.
It can be easily
understood with a very simple example that as $Re$ increases, the amplitude of vortices increases which are
indeed clear from the Figs. 7 and 8 given by Mukhopadhyay et al. (2005). 
Let us consider a 2D perturbation in
an inviscid incompressible flow where the vorticity $\nabla \times \vec{v}$ is exactly conserved,
when $\vec{v}=\hat{i} v_x+\hat{j} v_y$. Therefore, at $t=t_{max}=t_g$, when the perturbation
growth is maximum at $t=t_{max}$, the amplitude of vorticity 
$\zeta\sim|lv|\sim Re^{1/3}$. As $\nu_t$ and $\alpha$ are directly proportional to $\zeta^2$, they 
scale as $Re^{2/3}$ at $t=t_{max}=t_g$. Therefore, if $Re$ decreases three orders of magnitude, then
$\nu_t$ decreases in two orders. Moreover, the perturbation stabilizes at a thicker disk.
Indeed we find that the viscosity decreases, as $h(r)/r$ increases. Dimension of confined 
liquid in the experiments by Ji et al. (2006) may not be typical of astrophysical disks or 
rings, when they may have a large aspect ratio $\sim 2$, whereas the astrophysical 
disks and ring systems are normally thin (with aspect ratio $\leq 1$). 
Obviously a huge gap exists between experiments and the real observations.  

By numerical simulations, the formation and evolution of vortices in a hydrodynamic shearing-sheet
have already been studied by Johnson \& Gammie (2005) and they suggested it to be a possible mechanism for angular momentum transport 
in low-ionization disks at high resolution. It has been argued that there must be a mechanism
to inject vorticities into the disk, and the vortices must not decay rapidly due to threedimensional 
instabilities, to sustain the transport. We show that the vortices may sustain in 
threedimension at least in the time scale of interest, where this is applicable for accretion disks. 
Indeed, Cuzzi and his collaborators (\cite{cuzzi1,cuzzi2}) have argued, by
numerical simulations, that the elliptical instability may lead to turbulence to
from the dusty gas surrounding a young star. Also the vortex generation and
then the angular momentum transport has been
shown to occur in the unmagnetized protoplanetary disks (\cite{borro}) by hydrodynamic turbulence.
However, other simulations (\cite{ssg}) do not find significant transport.
The nonoccurence of significant transport in simulations,
in our view, is due to lack of resolution needed to capture the turbulence. Indeed, the later 
authors have mentioned that for their calculations it is difficult to define an effective Reynolds 
number, since the numerical dissipation is a steep function of resolution.
With a particular non-linear solution, Balbus \& Hawley (\cite{bh06}) have shown that 
perturbation decays asymptotically. They also have argued that as the nonlinear term in the
equation for the incompressible flow 
itself vanishes explicitly, the solution can not lead to nonlinearity and then
turbulence. However, this does not guaranty that every solution does so. 
They themselves have also mentioned that secondary
instabilities may still spoil their conclusion. 
Indeed the coupling between the secondary and primary modes was shown earlier not to allow the nonlinear
term to vanish resulting in a possible nonlinear transition to turbulence (\cite{mukh06}).

It is interesting to note that the modal instability via the bypass mechanism
(and then with a secondary perturbation superimposed) arises in these systems
from a subtle interplay of the non-normality
of the perturbation modes and the non-linearity of the Navier-Stokes equation and this in turn gives rise
to the turbulence in the system. As the turbulence and corresponding 
transport is inevitable in these systems, the
corresponding  $\alpha$ may not be just inversely proportional to the critical Reynolds number 
(as predicted earlier (\cite{ll})).
Previous theoretical studies (\cite{man}) have shown that the Keplerian flow may render
a transition to the turbulent regime at a Reynolds number
$\sim 10^{6}$ and turbulence might have just started at this critical
Reynolds number. It is to be seen now 
whether all shear flows, exhibiting subcritical turbulence in the laboratory,   
do exhibit large growth due to secondary perturbation.


\begin{acknowledgements}
This work is partly supported by a project, Grant No. SR/S2HEP12/2007, funded
by DST, India.
\end{acknowledgements}


\end{document}